\documentclass[aps,pra,nofootinbib,preprint,amsmath,amssymb,floatfix]{revtex4}
\usepackage{color}
\usepackage{graphicx}

\begin{document}

\newcommand{\tc}{\textcolor}
\newcommand{\g}{blue}
\newcommand{\ve}{\varepsilon}
\title{ Holographic cosmology with two coupled fluids  in the presence of viscosity}         

\author{  I. Brevik$^1$  }      
\affiliation{$^1$Department of Energy and Process Engineering,  Norwegian University of Science and Technology, N-7491 Trondheim, Norway}
\author{A. V. Timoshkin$^{2,3}$}
\affiliation{$^2$Tomsk State Pedagogical University, Kievskaja Street, 60, 634061 Tomsk, Russia}
\affiliation{$^3$International Laboratory of Theoretical Cosmology}
\affiliation{Tomsk State University of Control Systems and Radio Electronics, Lenin Avenue, 36, 634050 Tomsk, Russia}

\date{\today}          

\begin{abstract}
We explore the cosmological models of the late-time universe based on the holographic principle, taking into account the properties of the viscosity of the dark fluid. We use the mathematical formalism of generalized infrared cutoff holographic dark energy, as  presented by   Nojiri and  Odintsov (2017). We consider the Little Rip, the Pseudo Rip, and  a bounce exponential model, with two interacting fluids, namely  dark energy and dark matter  in a spatially-flat Friedmann-Robertson-Walker universe. Within these models, analytical expressions are obtained for infrared cutoffs  in terms of the particle horizons.  The law of conservation of energy is presented, from a holographic point of view.

\end{abstract}
\maketitle
Keywords: holographic principle; infrared cutoff; viscous dark fluid; equation of state.

\bigskip

\section{Introduction}

Application of the so-called   holographic principle \cite{1}  allows one to explain the accelerated expansion of the Universe.  An effective way to describe the evolution of the late Universe, is to make use of  the generalized cutoff holographic dark energy model of Nojiri and Odintsov \cite{2,3}.  Some analyses of this principle to explain the accelerated expansion of the Universe were carried out in Refs.~\cite{4,5,6,7,8,9,10,11,12,13}. Recently, the holographic principle has been applied to the inflationary Universe \cite{14,15},  as well as to  bouncing cosmology \cite{16,brevik20}.  It is known that the theory of holographic dark energy in the late-time Universe is well consistent with the data of astronomical observations  \cite{zhang05,li09,huang04,wang05}. Different applications of the theory of dark energy were studied in the reviews \cite{bamba12,nojiri17}.

The purpose of this article is to obtain a holographic description of the cosmology  of the late Universe, as well as the model with a rebound, with the help of the formalism of interacting fluids in the presence of viscosity. For the Little Rip, the Pseudo Rip and  a bounce exponential model, infrared radii are calculated in terms of the particle horizon  introduced by Nojiri and Odintsov \cite{3}. In the case of the late-time Universe, in addition to  holographic dark energy we consider also dark matter, unlike  the case of the early-time Universe where the material sector is not considered. As a result, a specific form of the law of conservation of energy in holographic form is obtained.

\section{Model  of  coupled dark fluids and holographic principle}

Let us consider a model of the Universe where there are  two interacting fluids, namely  dark energy and dark matter, in a spatially flat Friedmann-Robertson-Walker metric,
\begin{equation}
ds^2= -dt^2+a^2(t)\sum_{i=1}^3 (dx^i)^2. \label{1}
\end{equation}
We write the dynamic equations in the form \cite{24}
\begin{equation}
\dot{\rho}+3H(p+\rho)=-Q, \nonumber
\end{equation}
\begin{equation}
\dot{\rho}_m+3H(p_m+\rho_m)= Q, \label{2}
\end{equation}
\begin{equation}
\dot{H}=-\frac{1}{2}k^2(p+\rho +p_m +\rho_m), \nonumber
\end{equation}
 where $a$ is a scale factor, $H=\dot{a}/a$ is the Hubble parameter and $k^2=8\pi G$ is Einstein's gravitational constant with Newton's gravitational constant $G$;  $p,\rho$ and $p_m,\rho_m$ are the pressure and the energy density of dark energy and dark matter respectively. The term $Q$ in the right part of the equations describes the coupling between dark energy and dark matter. A dot denotes derivative with respect to the cosmic time $t$.

 The Friedmann equation for the Hubble function has the form \cite{24}
 \begin{equation}
 H^2= \frac{1}{3}k^2(\rho+\rho_m). \label{3}
 \end{equation}

The holographic model of dark energy is based on the holographic principle. We give the highlights following the terminology in Ref.~\cite{1}.  When the dark energy is described in this way it means that the horizon cutoff radius is related to the infrared cutoff.
According to the general holographic energy model \cite{3} the holographic  energy density is  inversely proportional to the square of the infrared cutoff $L_{IR}$,
 \begin{equation}
 \rho=  \frac{3c^2}{k^2L_{IR}^2}, \label{4}
 \end{equation}
 where $c$ is a dimensionless parameter.

Since there are no strong arguments about how to choose  an infrared radius $ L_ {IR} $, we select the  particle horizon  $ L_p $ or the event horizon $ L_f $ \cite {2}.   These are defined as
 \begin{equation}
 L_p= a\int_0^t\frac{dt}{a},  \quad L_f = a\int_t^\infty \frac{dt}{a}. \label{5}
\end{equation}

In the general case, the infrared cutoff $L_{IR}$ could be a combination of $L_p, L_f$ and their derivatives; it could contains also the Hubble function, the scale factor, and its derivatives \cite{3}. In the future, we will assume that the fluid driving the evolution of the universe has a holographic origin.

\section{Holographic representation of viscous fluid models}

 In this section we will explore the Little Rip, the Pseudo Rip and  a bounce exponential model. We assume that the viscous dark fluid interacting with dark matter is associated with holographic energy.

\subsection{Little Rip cosmology}

The characteristic feature of the Little Rip cosmology is that the energy density monotonously increases with time asymptotically. As a result, infinite time is required to achieve the  singularity. The  equation of state parameter is $\omega < -1$, but $\omega \rightarrow -1$ asymptotically. This is a soft variant of the singularity.

Let us consider the Little Rip model with the Hubble function \cite{25}
\begin{equation}
H= H_0 e^{\lambda t}, \quad H_0>0, \quad \lambda >0, \label{6}
\end{equation}
where $H_0=H(0)$,  $t=0$ is the present time.

Assuming  that dark matter is  dust, we have $p_m=0$.  Consequently, the law of conservation of energy for dark matter will take a simpler form
\begin{equation}
\dot{\rho}_m+3H\rho_m=Q. \label{7}
\end{equation}
We will use the same coupling as in Ref.~\cite{nojiri11}
\begin{equation}
Q= \delta H\rho_m, \label{8}
\end{equation}
where $\delta$ is a positive nondimensional constant. In view  of Eq.~(\ref{8}) the solution of Eq.~(\ref{7}) is \cite{27}
\begin{equation}
\rho_m(t)=\rho_0\exp \left( \frac{\delta-3}{\lambda}H\right),\label{9}
\end{equation}
where $\rho_0$ is an integration constant.

Let us consider the following  equation of state for the viscous fluid \cite{28}:
\begin{equation}
p= \omega(\rho,t)\rho -3H\zeta(H,t), \label{10}
\end{equation}
where $\omega(\rho,t)$ is a thermodynamic parameter and $\zeta(H,t)$ is the bulk viscosity, which depends on the Hubble function and on the time $t$. From thermodynamic considerations it follows that $\zeta(H,t)>0$.

Let us consider the simplest case, when the thermodynamic parameter  $\omega(\rho,t)=\omega_0$ and the bulk viscosity $\zeta(H,t)= \zeta_0$ are constants. Then Eq.~(\ref{10}) will take the form
\begin{equation}
p=\omega_0\rho-3\zeta_0H. \label{11}
\end{equation}
Let us calculate the scale factor
\begin{equation}
a(t)= a_0\exp \left( \frac{H_0}{\lambda}\exp (\lambda t)\right) \label{12}
\end{equation}
and the particle horizon $L_p$
\begin{equation}
L_p = \frac{1}{\lambda}\exp\left( \frac{1}{\lambda}H\right) \left[Ei\left( -\frac{1}{\lambda}H\right)-Ei\left( -\frac{1}{\lambda}H_0\right)\right], \label{13}
\end{equation}
where $Ei(bx), \, b \neq 0$ is the integral exponential function \cite{29}.

In the holographic language, the Hubble function can be expressed through the particle horizon $L_p$ \cite{3},
\begin{equation}
H= \frac{\dot{L}_p-1}{L_p}, \quad \dot{H}= \frac{\ddot{L}_p}{L_p}- \frac{\dot{L}_p^2}{L_p^2}+\frac{\dot{L}_p}{L_p^2}. \label{14}
\end{equation}
Let us write the corresponding representation of the energy conservation equation of the Little Rip model
\begin{equation}
(\omega_0+1)\left( \frac{\dot{L}_p-1}{L_p}\right)^2 =
\frac{1}{3}\omega_0\rho_0k^2\exp\left[ \frac{2(\delta-3)}{3\zeta_0k^2}\,
\frac{\dot{L}_p-1}{L_p}\right], \label{15}
\end{equation}
assuming that $\lambda=\frac{3}{2}\zeta_0k^2$.

Thus, we have obtained a reconstruction of the conservation equation as a generalized form of holographic energy.

\subsection{Pseudo Rip cosmology}

We will study a Pseudo Rip model with the Hubble function \cite{25}
\begin{equation}
H= H_1-H_0\exp(-\tilde{\lambda}t), \label{16}
\end{equation}
where $H_0, H_1$ and $\tilde{\lambda}$ are positive constants, $H_0>H_1,\, t>0$. In the early-time Universe, when $t\rightarrow 0$ we have $H \rightarrow H_1-H_0$, while  in the late-time universe, when $t\rightarrow +\infty$ the Hubble function tends to a cosmological constant, $H \rightarrow H_1$.

In this case, taking into account (\ref{8}), we obtain the solution of Eq.~(\ref{7}) \cite{27}
\begin{equation}
\rho_m(t)= \rho_0\exp \left[ (\delta-3)\left( H_0 t-
\frac{H-H_0}{\tilde{\lambda}}\right)\right], \label{17}
\end{equation}
where $\rho_0$ is an integration constant.

Now, assume $\omega(\rho,t)=\omega_0$ as before, and take the bulk viscosity to be proportional
to the Hubble function, $\zeta(H,t)=3\tau H$, where  $\tau$  is a positive-dimensional constant.

The equation of state (\ref{10}) takes the form
\begin{equation}
p= \omega_0\rho -9\tau H^2. \label{18}
\end{equation}

We can calculate the scale factor
\begin{equation}
a(t)= a_0\exp\left[ H_0t+\frac{H_1}{\tilde{\lambda}}\exp(-\tilde{\lambda}
t)\right]. \label{19}
\end{equation}
Let us put $\tilde{\lambda}=H_0$ and calculate the particle horizon $L_p$:
\begin{equation}
L_p= \frac{1}{H_1}\left\{ 1-\exp\left[ -\frac{H_1}{H_0}(1-\exp(-H_0t))\right]\right\}\exp(H_0t). \label{20}
\end{equation}

 In this case, the law of conservation of energy in holographic form becomes

\begin{equation}
2\left( \frac{\ddot{L}_p}{L_p}-\frac{\dot{L}_p^2}{L_p^2}+ \frac{\dot{L}_p}{L_p^2}\right) +
3(\omega_0-3\tau k^2+1)\left( \frac{\dot{L}_p-1}{L_p}\right)^2 = \omega_0k^2\rho_m. \label{21}
\end{equation}

Thus, a holographic representation of the Pseudo Rip model for a viscous dark fluid interacting with dark matter was obtained.

\subsection{Bounce cosmology}

In this section we consider an example of cyclic cosmology or cosmology with rebound \cite{30,31,32}. This name is due to the fact that at the initial instant of time, the Universe, filled with matter, is in a compressed state. Then there is a rebound without the formation of singularity, after which the accelerated expansion begins, corresponding to the inflationary stage.

 Let us suppose  that the bounce is described by a viscous fluid having a holographic origin.
 We again obtain the energy conservation law in terms of the  particle horizon. Note  that the coupling of a viscous fluid with dark matter allows one to achieve a better agreement of the theoretical models with  astronomical observations  \cite{28}.

 Let us consider a bounce exponential model \cite{33},
 \begin{equation}
 a(t)= a_0\exp[\beta(t-t_0)^{2n}], \label{22}
 \end{equation}
 where $a_0, \beta$ are positive dimensional constants, $n \in N$, and $t_0$ a fixed bounce time.

The evolution of the universe in this model is as follows.  Before bounce time $(t<t_0)$, the scale factor decreases and there occurs
 a contraction of the Universe. At the  instant  $(t=t_0)$, when $a_0= a(t_0)$, the bounce takes place. Later, at  $(t>t_0)$, the scale factor increases and the Universe expands.

The Hubble function is given by the expression
 \begin{equation}
 H(t)= 2n\beta (t-t_0)^{2n-1}. \label{23}
 \end{equation}

Let us suppose that the coupling function $Q(t)$ has the same form (\ref{8}) as above. Then the matter energy density equals \cite{27}
 \begin{equation}
 \rho_m(t)= \rho_0\exp\left[ \frac{\delta-3}{2n}H(t)(t-t_0)\right], \label{24}
 \end{equation}
 where $\rho_0$ is an integration constant.

Let us consider the case with $\zeta(H,t)= \zeta_0$ and  a linear dependence of the thermodynamic parameter with time,
 \begin{equation}
 \omega(\rho,t)=a_1t+b_1, \label{25}
 \end{equation}
 where $a_1,b_1$ are arbitrary constants.

 Then the equation of state (\ref{10}), is obtained as follows
 \begin{equation}
 p=(a_1t+b_1)\rho-3\zeta_0H. \label{26}
 \end{equation}

 Let us calculate the particle horizon
 \begin{equation}
 L_p= \frac{1}{2}\sqrt{\frac{\pi}{\beta}}\exp[ \beta (t-t_0)^2]{\rm erf}(\sqrt{\beta}\, t), \label{27}
 \end{equation}
 where erf$(\sqrt{\beta}\, t)$ is the probability integral.

Finally, we write the  energy conservation law as
 \begin{equation}
 2\left( \frac{\ddot{L}_p}{L_p}-\frac{\dot{L}_p^2}{L_p^2}+ \frac{\dot{L}_p}{L_p^2}\right) +
 9(a_1t+b_1+1)\left( \frac{\dot{L}_p-1}{L_p}\right)^2-9\zeta_0k^2\left( \frac{\dot{L}_p-1}{L_p}\right)=
 3(a_1t+b_1)k^2\rho_m. \label{28}
\end{equation}
 Thus, we have obtained a holographic representation of the bounce exponential model for a viscous fluid coupled with matter.

 \section{Conclusion}

This work  is devoted to the holographic description of the cosmological models Little Rip, Pseudo Rip, and the bounce exponential model,  in a homogeneous and isotropic
Friedmann-Robertson-Walker metric. In order to apply the holographic principle to these models, we have identified the infrared radius $L_{IR}$ with the  particle horizon  $L_p$. As a model of the Universe, we have assumed a dark viscous fluid interacting with dark matter. We have applied the holographic principle for cosmological models for various values of the thermodynamic
 parameter $\omega(\rho,t)$ and for different forms of the bulk viscosity $\zeta(H,t)$ in the equation of state. In these models, the infrared radius in the form of a particle horizon have been  calculated, in order to obtain the appropriate energy conservation law in each case.  In that way, we have established the equivalence between the viscous models and the holographic models.

The natural question arises: is the holographic theory in agreement with astronomical observations? A comparative analysis, involving  a theoretical holographic model of dark energy on a brane, was given  in \cite{34}. It  showed that,  for a wide range of parameters including redshifts  for distant supernova Ia, good agreements were obtained between  observed data and theoretical predictions.

\section*{Acknowledgement}

 This work was supported in part by Ministry of Education of Russian Federation, Project No FEWF-2020-0003 (A.V.T.).

\end{document}